\documentclass[twocolumn]{aastex631}

\usepackage{multirow}

\newcommand{\kms}{km s$^{-1}$}
\def \iris   	{{\it IRIS}}
\def \alfven  {Alfv$\acute{\rm e}$n}
\def \siiv {\ion{Si}{4}}
\def \mgii {\ion{Mg}{2}}
\def \fexviii {\ion{Fe}{18}}
\def \fexii {\ion{Fe}{12}}
\def \oiv {\ion{O}{4}}

\begin{document}

\title{A Statistical Study of \iris\ Observational Signatures of Nanoflares and Non-thermal Particles}

\author[0000-0001-7460-725X]{Kyuhyoun Cho}
\affiliation{Bay Area Environmental Research Institute, NASA Research Park, Moffett Field, CA 94035, USA}
\affiliation{Lockheed Martin Solar \& Astrophysics Laboratory, 3251 Hanover Street, Palo Alto, CA 94304, USA}

\author[0000-0002-0405-0668]{Paola Testa}
\affiliation{Harvard-Smithsonian Center for Astrophysics, 60 Garden Street, Cambridge, MA 02193, USA}

\author[0000-0002-8370-952X]{Bart De Pontieu}
\affiliation{Lockheed Martin Solar \& Astrophysics Laboratory, 3251 Hanover Street, Palo Alto, CA 94304, USA}
\affiliation{Rosseland Centre for Solar Physics, University of Oslo, P.O. Box 1029 Blindern, NO-0315 Oslo, Norway}
\affiliation{Institute of Theoretical Astrophysics, University of Oslo, P.O. Box 1029 Blindern, NO-0315 Oslo, Norway}

\author[0000-0002-4980-7126]{Vanessa Polito}
\affiliation{Bay Area Environmental Research Institute, NASA Research Park, Moffett Field, CA 94035, USA}
\affiliation{Lockheed Martin Solar \& Astrophysics Laboratory, 3251 Hanover Street, Palo Alto, CA 94304, USA}
\affiliation{Department of Physics, Oregon State University, 301 Weniger Hall, Corvallis, OR 97331}



\begin{abstract}
Nanoflares are regarded as one of the major mechanisms of magnetic energy release and coronal heating in the solar outer atmosphere. We conduct a statistical study on the response of the chromosphere and transition region to nanoflares, as observed by the {\it Interface Region Imaging Spectrograph} (IRIS), by using an algorithm for the automatic detection of these events. The initial atmospheric response to these small heating events is observed, with \iris, as transient brightening at the footpoints of coronal loops heated to high temperatures ($>4$MK). For four active regions, observed over $143$~hours, we detected 1082 footpoint brightenings under the \iris\ slit, and for those we extracted physical parameters from the \iris\ \mgii\ and \siiv\ spectra that are formed in the chromosphere and transition region, respectively. We investigate the distribution of the spectral parameters, and the relationship between the parameters, also comparing them with predictions from RADYN numerical simulations of nanoflare-heated loops. We find that these events, and the presence of non-thermal particles, tend to be more frequent in flare productive active regions, and where the hot Atmospheric Imaging Assembly 94 \AA\ emission is higher. We find evidence for highly dynamic motions characterized by strong \siiv\ non-thermal velocity (not dependent on the heliocentric $x$ coordinate, i.e., on the angle between the magnetic field and the line-of-sight) and asymmetric \mgii\ spectra.  These findings provide tight new constraints on the properties of nanoflares, and non-thermal particles, in active regions, and their effects on the lower atmosphere.
\end{abstract}



\section{Introduction} \label{sec:intro}
Nanoflares, along with magnetohydrodynamic (MHD) \alfven\ waves, are thought to be an important mechanism to explain the heating of the solar corona, which,  despite being of fundamental importance in astrophysics, is still not well understood \citep[e.g.,][]{Klimchuk2006,Testa2015,Testa_Reale_2022arXiv}. Nanoflares can be produced by magnetic reconnection via braiding of coronal magnetic field lines caused by photospheric motions \citep{Parker1988}. The effect of these small energy releases (thought to be of the order of $10^{24}$-$10^{25}$~erg) is often difficult to observe in the corona because of the high-conductivity in the corona efficiently spreading out the energy, and the weak initial emission due to the low emission measure as well as non-equilibrium processes. The lower atmosphere of the coronal loops is more sensitive to the heating release in the initial phases and it provides useful diagnostics of coronal heating \citep[e.g.,][]{Testa2013,Testa2014}. 

In the standard model of the solar flares, relativistic particles, which are accelerated by magnetic reconnection in the corona, penetrate along the magnetic fields and deposit energy in the chromosphere via collisions \cite[e.g.,][]{Holman2011}. Similarly, nanoflares are also expected to generate brightenings in the chromosphere and transition region, and, if indeed they are scaled-down version of larger flares, they might also show presence of accelerated particles.
However, these non-thermal electron beams in nanoflares are likely characterized by shorter duration and smaller energies than in larger flares.
Indeed, when non-thermal particles are observed in smaller heating events (nanoflare to microflares), their distributions are typically characterized by smaller low-energy cutoff and steeper slopes ($E_{\rm C} \sim 5-15$~keV, $\delta \gtrsim 7$; \citealt{Hannah2008,Testa2014,Wright2017,Testa2020,Glesener2020,Cooper2021}) than larger flares, and therefore also penetrating less deep in the low atmosphere.

Recently, small scale transient brightenings have been reported as the signature of nanoflares in high resolution and high cadence observations. \cite{Testa2013} reported that the footpoints of hot loops, known as moss, occasionally show high variability in Hi-C 193 \AA\ data. These footpoint brightenings are characterized by timescales of the order of $\sim$15~s, which is much shorter than the known time scale of the typically observed moss variability \citep[e.g.,][]{DePontieu1999,Antiochos2003}. Thus, smaller energies and shorter timescales, compared with larger flares, are required to explain these short term brightenings. The {\it Interface Region Imaging Spectrograph} (IRIS; \citealt{DePontieu2014}) provided valuable clues on the heating mechanisms of transient hot loops \citep{Testa2014,Testa2020}. \iris\ provides high temporal and spatial resolution UV spectra formed in the chromosphere and transition region. 
The comparison of the observed temporal evolution of spectral properties of \iris\ lines with the prediction of numerical simulations of nanoflare heated loops indicated that the heating by non-thermal electrons is more plausible than conduction to explain some of the observations \citep{Testa2014,Testa2020}.
\cite{Polito2018} and \cite{Testa2020} performed RADYN simulations to show the effect of several physical parameters of the nanoflares (e.g., duration, total energy, low-energy cutoff of non-thermal electron distribution), as well as the initial conditions of the loops, on the observable spectra. They showed that, when non-thermal electrons are present, the energy deposition layer in the lower atmosphere will be determined by the density of coronal loops and the hardness and total energy of the electron beam. Consequently, the properties of the heating also determine the sign of the Doppler velocities, and the intensities of the chromospheric and transition region lines. \cite{Bakke2022} recently used the same RADYN simulations to investigate additional chromospheric diagnostics of the heating properties based on ground-based spectral observations. These numerical models provide a useful framework to interpret the observed spectra. 

The recent \iris\ observational studies of these footpoint brightenings have analyzed a limited number (about a dozen; \citealt{Testa2014,Testa2020}) of events, manually selected, precluding any definitive conclusion on the general properties of nanoflares, and presence and properties of non-thermal particles, in active regions, outside large flares. 
In order to overcome these limitations we performed statistical studies of properties of nanoflare events, by using an automated selection procedure to identify a large sample of events, and therefore also reducing some selection biases possibly present in previous studies. For this purpose, we exploit the several EUV images observed by {\it Atmospheric Imaging Assembly} (AIA; \citealt{Lemen2012}) on board the {\it Solar Dynamics Observatory} (SDO; \citealt{Pesnell2012}) to detect the moss location, and investigate the \iris\ spectra obtained at those locations for four different active regions. 
We apply a modified version of the automated algorithm of \citet{Graham2019} to automatically detect footpoint variability in co-aligned AIA-\iris\ datasets of several active regions, and to extract the \iris\ spectral line properties (intensity, Doppler shift, broadening) of \mgii\ and \siiv\ for all transient brightenings caught under the \iris\ slit. 
Through the analysis of the  statistical distribution of the physical quantities, and of the correlations between different parameters, we determined the general characteristics of nanoflares and their effects on the lower atmosphere. We also compare our finding with previous observational studies and numerical simulation results, and discuss the effect of nanoflares on the lower layer in terms on the plasma dynamics and heating mechanism.

\section{Observation and Analysis} \label{sec:analysis}
We searched the \iris\ database for observations suitable to detect transient loop footpoint brightenings and their chromospheric and transition region emission. We limited the search to sit-and-stare mode, including the \siiv\ 1402.77 \AA\ and \mgii\ h\&k spectral windows, with raster scan cadence of less than 15~s to ensure high cadence for the spectral observations.  We searched for \iris\ datasets including several observations of the same active region over several days, so we can also investigate the dependence of the observed heating properties on line-of-sight, and on the active region activity level.  
Using \texttt{ssw\_hcr\_query.pro} in IDL solarsoft, we searched the \iris\ observing time for every active region from active region number 11800 to 13050 on the solar disk. It is possible that our approach means that data before the assignment of an active region number might not be included. In addition, only \iris\ data coincident in time with relevant AIA data without gaps due to missing data or eclipses were selected. Here we present the analysis of \iris\ observations for four suitable active regions. 

As in \citet{Graham2019}, the small scale transient brightening moss regions are identified in AIA data. We used the co-aligned AIA datacubes produced for each \iris\ observation, and available on the \iris\ search page\footnote{\url{https://iris.lmsal.com/search/}}. After exposure time normalization, we applied the rapid varying moss detection algorithm \citep{Graham2019} to the co-aligned AIA data. Here we briefly describe the algorithm. We identify the network region with  AIA 1700 \AA\ emission above a threshold ($>$80~DN). Then, we select datasets where hot loops are present because we want to investigate the moss variability related to coronal heating events producing hot loops. The \fexviii\ emission in the AIA 94 \AA\ passband traces plasma hotter than $\sim $4~MK. In particular, to separate the \fexviii\ emission from the other cooler contributions to the 94 \AA\ emission we employed the following simple relation using three different AIA channel data \citep{DelZanna2013}:
\begin{equation}
I_{\textrm{\tiny{\fexviii}}} = I_{94}-\frac{I_{211}}{120}-\frac{I_{171}}{450}.
\end{equation}
We chose the pixels where \fexviii\ emission is greater than 5~DN, and the size of the coronal hot loop is larger than 100~pixels which is an empirical value for the minimum size of a coronal loop. To select bright moss emission, we identify the regions with AIA 193 \AA\ intensity greater than 1250~DN.
Even after all these criteria have been applied, it is possible that the selected areas might be associated with other phenomena such as filament eruptions or large flares. To eliminate these possibilities, we used the differential emission measure method (DEM, \citealt{Cheung2015}): we excluded areas where the emission measure in two temperature bands $\log T = 5.6$ - $5.8 $ and $\log T = 6.7$ - $7.0$ is greater than $2 \times 10^{26} $ cm$^{-5}$ and $2.5 \times 10^{27} $ cm$^{-5}$, respectively, and their area is less than 15 pixels. We manually confirmed that this method effectively removes flares and filament eruption in most cases. Because we are interested in transient brightening events, we investigated the temporal variation in 171 \AA\ and 193 \AA\ and selected the local maximum in the light curve. From the first derivative of the light curve, we selected the times when its sign changed from positive to negative, known as zero-crossings. \citep[e.g.,][]{Testa2013,Graham2019}.

After identifying the position and time of the moss brightenings in the AIA data, we checked the corresponding \iris\ slit position and observing time. If the \iris\ dataset includes the 1400 \AA\ slit-jaw images (SJIs), we repeated the alignment process of these with the AIA 1600 \AA\ images to increase the accuracy of the \iris-AIA alignment. We collected every IRIS raster pixel which is corresponding spatially and temporally, within 1 arcsecond and 6 seconds, to the selected AIA moss position and time. We set an additional criterion using the \siiv\ spectral line. We produce the \siiv\ light curves deriving the temporal evolution of the \siiv\ total intensity obtained by integrating over 1402.77$\pm2$ \AA. The light curves obtained for an interval of $\pm 150$ seconds from the moss detection instant in each pixel. Then, we investigated if an intensity peak exists within $\pm 30$ seconds from the moss detecting instant, and the intensity peak value is greater than 3$\sigma$ of light curve from the base intensity. We also checked that the intensity peak has a lifetime (FWHM) shorter than 60 seconds. If the \siiv\ light curve satisfies these three conditions the event is marked as a moss brightening and included in our sample. Then, using the chi-square values from the spectral fitting, as described later, we discard the bottom 1\%, which effectively eliminates poorly fitted spectra. Finally, we conducted the investigation on the spectral properties at the peak time of the \siiv\ light curve in each pixel. For the \mgii\ spectral analysis, we used the peak time of \mgii\ k light curve in each pixel obtained by integrating within $\pm0.65$ \AA\ of its central wavelength. As a result, a total of 1082 pixels were obtained from 747 moss brightening events, and only the peak time properties were taken from each pixel. The information and number of selected pixels for each active region is summarized in Table \ref{AR_info}. 

\begin{deluxetable*}{ccccc}[t]
\tablecaption{Information about targeted active regions \label{AR_info}}
\tablenum{1}
\tablewidth{0pt}
\tablehead{
\colhead{\begin{tabular}[c]{@{}c@{}}Active region\\ number\end{tabular}} & 
\colhead{\begin{tabular}[c]{@{}c@{}}Observing \\ times (hr)\end{tabular}} & 
\colhead{\begin{tabular}[c]{@{}c@{}}Selected number \\ of moss pixels\end{tabular}} & 
\colhead{\begin{tabular}[c]{@{}c@{}}Number of flares\tablenotemark{a} \\ (C / M / X) \end{tabular}}  &
\colhead{\begin{tabular}[c]{@{}c@{}}The Mount Wilson \tablenotemark{a, b} \\ magnetic classification \end{tabular}}}  
\startdata
12415 & 47.98 & 362 &  36 / 2 / 0 & $\beta  \; / \; \beta\gamma$  \\
12473 & 43.63 & 439 &  30 / 5 / 0 & $\beta  \; / \; \beta\gamma \; / \; \beta\delta \; / \; \beta\gamma\delta$ \\
12524 & 11.09 &  36 &   3 / 0 / 0 & $\alpha \; / \; \beta$   \\
12529 & 40.28 & 245 &  30 / 1 / 0 & $\alpha \; / \; \beta \; / \; \beta\gamma$ \\
\hline
Total & 142.98 & 1082 &           \\
\enddata
\tablenotetext{a}{\url{http://helio.mssl.ucl.ac.uk/helio-vo/solar_activity/arstats-archive}}
\tablenotetext{b}{\citet{Hale1919, Kunzel1965}}
\end{deluxetable*}

We collect the spatial information for the \iris\ moss brightenings. The heliocentric coordinate of the pixels provides information about the inclination of the local vertical with respect to the line of sight. Under the reasonable assumption that, on average, the magnetic field in the transition region footpoints of hot active region loops is typically vertical, this then provides a constraint on dependence with respect to the magnetic field direction. The spatially averaged \fexviii\ emission for whole data field of view (FOV) provided with \iris\ data was collected as an auxiliary value for the environment of the event. 

We extract the spectral parameters by fitting the \iris\ spectra normalized by their exposure time. For the \siiv\ spectra, we use a single Gaussian function to fit the 1402.77$\pm2$ \AA\ wavelength range and obtain the three parameters -- amplitude, centroid, and width -- from which we derive the intensity amplitude, Doppler velocity, and non-thermal velocity, respectively. The non-thermal velocity was acquired by subtracting the thermal broadening and the instrumental broadening as follows 
\begin{equation}
w_{nth} = \sqrt{w_{obs}^2 - w_{th}^2 - w_{I}^2}
\end{equation}
where $w_{nth}$ is the non-thermal velocity, $w_{obs}$ is the $1/e$ width of \siiv\ line which is corresponds to $\sqrt{2}$ times the Gaussian width, $w_{th}$ is the thermal velocity which corresponds to about 6.9 km s$^{-1}$ for the Si {\sc iv} 1402 \AA, and $w_{I}$ is the instrumental broadening which is order of 3.3 km s$^{-1}$ for the \iris\ FUV spectral band \citep{DePontieu2014}. 

We also calculated the plasma electron density using the diagnostic based on the \oiv\ 1399 \AA\ and 1401 \AA\ line ratio\footnote{\url{https://iris.lmsal.com/itn38/diagnostics.html##density-diagnostics}}. We obtained the \oiv\ total line intensities by integrating between 1399.77$\pm$0.25 \AA, and 1401.16$\pm$0.25 \AA\ respectively, and then calculated the line ratio. We exclude the samples of which total line intensity is lower than 3 sigma. The sigma is defined by following equation:
\begin{equation}
\sigma_{line} = \sqrt{N_{pix} \sigma_{bg}^2}
\end{equation}
where $\sigma_{line}$ is sigma for the total line intensity, $N_{pix}$ is the number of spectral pixels for each \oiv\ line, and $\sigma_{bg}$ is the standard deviation of background spectra in the wavelength range from 1404.25$\pm$0.25 \AA, where no noticeable spectral line exists \citep{Curdt2004}. We collect the density information in 55\% (593/1082) of the pixels. The others have the line ratio outside of the theoretically expected range between 0.17 and 0.42 for density diagnostics, or fail the 3 sigma detection criterion. These  two groups have a large overlap. 

For the \mgii\ h\&k lines, we measured the velocity using two different methods. First, we calculate the centroid of the spectra using the center of gravity method in the range within $\pm 0.65$ \AA\ of their central wavelengths, and convert these values to velocities, 
\begin{equation}
v_{center} = \frac{c}{\lambda_0} \frac{\int (\lambda-\lambda_0) I_{\lambda} \, d\lambda }{\int I_{\lambda}
\, d\lambda}
\end{equation}
where $v_{center}$ is the centroid velocity, $c$ is speed of light, $\lambda$ is wavelength, $\lambda_0$ is central wavelength of the lines, and $I_{\lambda}$ is observed spectrum. This centroid velocity roughly reflects the average motion within the chromosphere because the formation height of the \mgii\ h\&k lines spans a relatively wide height range. Second, we measure the \mgii\ h3 and k3\footnote{See Figure 3.2 in \url{https://iris.lmsal.com/itn39/Mg_diagnostics.html##the-mg-ii-h-k-lines}} Doppler velocities. As in \cite{Schmit2015}, we adopted a combination of two symmetric linear functions, and a positive and a negative Gaussian functions to fit each line. This function is coded as \texttt{iris\_mgfit.pro} in IDL Solarsoft. We determined k3 and h3 positions from the fitted spectral curve through \texttt{iris\_postfit\_get.pro} which is also included in the IDL Solarsoft, then these spectral positions were converted to the velocities. If the measured h3 and k3 velocities well represent the dynamics at $\tau=1$ layer like in numerical simulations of the quiet Sun \citep{Leenaarts2013}, they can provide valuable information about the upper chromospheric dynamics. However, the findings by \citet{Leenaarts2013} might not necessarily apply to the very dynamic active region plasma we are investigating here. Futhermore, in some cases the observed \mgii\ spectra are either single peaked, or with multiple ($>$2) peaks, so it is not always easy to determine the h3 and k3 positions. In fact, about 19\% of \mgii\ k lines (206/1082) and \mgii\ h lines (203/1082) show single peak profiles, and about 16\% cases (107/1082) show single peak profiles in both spectral lines. The single peaks are often accompanied by highly shifted tiny k3 or h3 self absorption features. To avoid misidentifications, we excluded those from the \mgii\ parameters analysis. On the contrary, the centroid method is free from this problem, so we mainly use centroid velocities to investigate chromospheric dynamics, {and the h3 and k3 velocities are used as supplementary parameters. As we will discuss later, another advantage of the centroid analysis is that the comparison with the RADYN simulations \citep{Polito2018,Testa2020} is more straighforward as the synthetic \mgii\ profiles are often quite complex, rendering determination of the k3 and h3 Doppler shifts difficult. 

Another important feature is the \mgii\ triplet line at 2798.823 \AA. Numerical simulations in \cite{Testa2020} showed that this line is expected to be observed as an emission line when non-thermal electron of sufficient energy ($\gtrsim 10$~keV) are present, as they will deposit energy and locally heat the lower chromospheric layer. In general, the \mgii\ triplet spectrum exhibits a complicate shape which is like a miniature of \mgii\ h\&k line: it has dips at both wings and self absorption at the core similar to the \mgii\ h1\&k1 or h3\&k3, so it is not easy to determine whether it is an emission line or not. To quantify its behavior, we calculated the equivalent width of the \mgii\ triplet line by integration of spectrum over 2798.823$\pm$0.2 \AA\ interval
\begin{equation}
\textrm{EW} = \int \frac{I_{\lambda}-I_{c}}{I_{c}} \, d\lambda
\end{equation}
where EW is equivalent width, and $I_{c}$ is the continuum determined by quadrature fitting within 2797.5 \AA\ to 2802.5 \AA\ range. We defined the \mgii\ triplet to be in emission if its equivalent width is a positive value. 

For the comparison with numerical simulations, we adopted the models from \cite{Polito2018} and \cite{Testa2020}, which are produced by using the RADYN code (\citealt{Carlsson1992, Carlsson1995, Carlsson1997a, Allred2015}). That is a 1-dimensional hydrodynamic code including non-local thermodynamic equilibrium radiative transfer. We obtain the optically thin \siiv\ synthetic spectra as described in \cite{Testa2014}, \cite{Polito2018} and \cite{Testa2020}. The \mgii\ spectra are calculated by using the RH 1.5D radiative transfer code (\citealt{Uitenbroek2001, PereiraUitenbroek2015}) based on the atmospheric model from the RADYN simulations. The detailed information of the simulation models is summarized in Table \ref{sim_model}. All models considered here assume the same initial atmosphere characterized by $\sim 1$~MK loop top temperature, coronal density of $\sim 5 \times 10^8$~cm$^{-3}$, and loop length of 15~Mm (see \citealt{Polito2018} and \citealt{Testa2020} for more details). We analyzed the synthetic \iris\ \mgii\ and \siiv\ spectral profiles at the peak time of their light curve using the same procedure applied to the observed spectra. This allowed us to derive spectral parameters of synthetic spectra that can be directly compared with ones derived from the observed spectra. We note that different models are characterized by different peak times for the transition region and chromospheric emission.

\begin{deluxetable*}{lccccc}[t]
\tablecaption{Parameters of RADYN simulations of nanoflare heated loops \label{sim_model}}
\tablenum{2}
\tablewidth{0pt}
\tablehead{
\colhead{\begin{tabular}[c]{@{}c@{}}Model\tablenotemark{a} \\ \, \end{tabular}} & 
\colhead{\begin{tabular}[c]{@{}c@{}}Total energy \\ (10$^{24}$ erg) \end{tabular}} & 
\colhead{\begin{tabular}[c]{@{}c@{}}Flux \\ (10$^9$ erg s$^{-1}$ cm$^{-2}$) \end{tabular}} & 
\colhead{\begin{tabular}[c]{@{}c@{}}Cutoff energy \\ (keV) \end{tabular}} & 
\colhead{\begin{tabular}[c]{@{}c@{}}Spectral index \\ \, \end{tabular}} & 
\colhead{\begin{tabular}[c]{@{}c@{}}Duration \\ (s) \end{tabular}}
}
\startdata
C1  &  6 & N/A & N/A & N/A & 10 \\
C1+ & 10 & N/A & N/A & N/A & 10 \\
E1  &  6 & 1.2 &  5  &  7  & 10 \\
E1+ & 10 & 2.0 &  5  &  7  & 10 \\
E2  &  6 & 1.2 & 10  &  7  & 10 \\
E2+ & 10 & 2.0 & 10  &  7  & 10 \\
E3  &  6 & 1.2 & 15  &  7  & 10 \\
E3+ & 10 & 2.0 & 15  &  7  & 10 \\
E4  &  6 & 0.4 & 10  &  7  & 30 \\
E5  & 18 & 1.2 & 15  &  7  & 30 \\
E6  &  6 & 0.6 & 10  &  7  & 20+60\tablenotemark{b} \\
H1  &  6 & 0.6 & 10  &  7  & 10 \\
H2  & 18 & 0.6 & 10  &  7  & 30 \\
\enddata
\tablenotetext{a}{C and E indicate heating by thermal conduction, and heating by accelerated electrons, respectively. H indicates hybrid models which include heating by both local coronal heating and thermal conduction, and accelerated electrons. + indicates higher total energy (10$^{25}$ erg) model. The model labels follow the naming convention used in \citet{Testa2020}.}
\tablenotetext{b}{intermittent heating: 20~s heating, 60~s pause, and 20~s heating again}
\end{deluxetable*}

\section{Results} \label{sec:results}

\begin{figure*}
\plotone{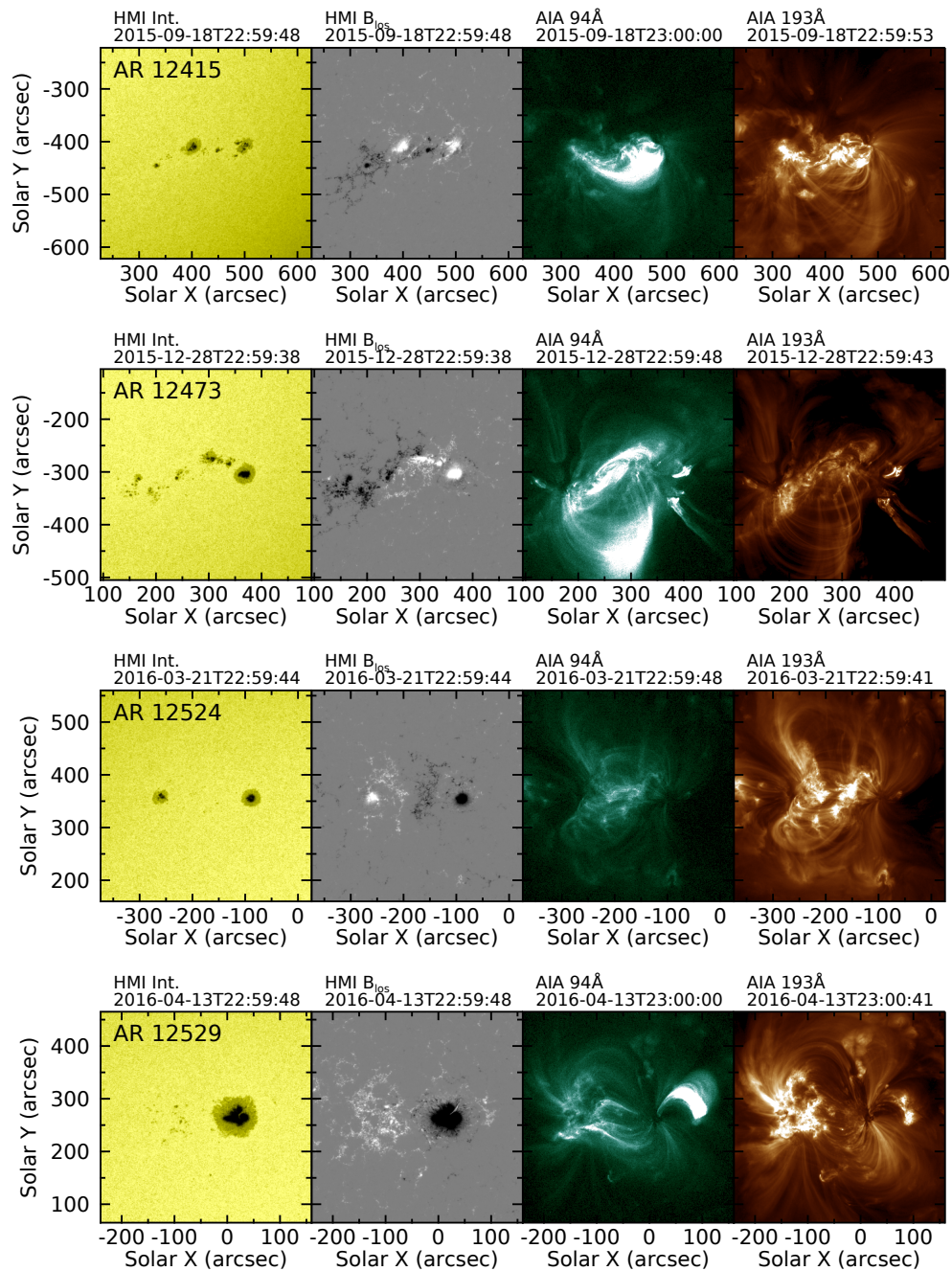}
\caption{Targeted active region images. Each column shows SDO/HMI continuum intensity, line of sight magnetogram, AIA 94 \AA, and 193 \AA\ data. Each row indicates different active region.}
\label{ar_image}
\end{figure*}

Figure \ref{ar_image} shows SDO/AIA observations of the four selected target active regions, and in particular, we show images in the photospheric intensity, line of sight magnetic field maps, 94 \AA\ (in active region cores typically dominated by \fexviii\ emission; \citealt{Testa2012b,Cheung2015}), and 193 \AA\ images (where active region moss is bright). Even though the active regions evolved as they crossed the solar disk during their life time, their characteristics are reflected in these images. For example, in the case of active region 12524, it shows a relatively simple configuration of the magnetic polarities, and it is well reflected in their Mount Wilson magnetic classification in Table \ref{AR_info}. It shows dimmed 94 \AA\ intensity compared to the other active regions. Throughout the observed period this active region was less active than the other active regions. Here we are interested in identifying and characterizing small heating events and exclude general flares, and we filter them out also using the DEM analysis, as described in the previous section. 
The number of observed flare events in each active region is summarized in Table~\ref{AR_info}. We found more small-scale transient brightening pixels in active regions where flaring activity is strong and the magnetic field configuration is complex. It is a well known fact that the flare activity of active regions is strongly correlated with their Mount Wilson magnetic classification \citep[and reference therein]{Toriumi2019}. Although the probability for observing the brightening pixels is affected by the slit location and observing time, it appears that the small scale-transient brightenings have a relation to the magnetic field configuration as well.

\begin{figure*}
\plotone{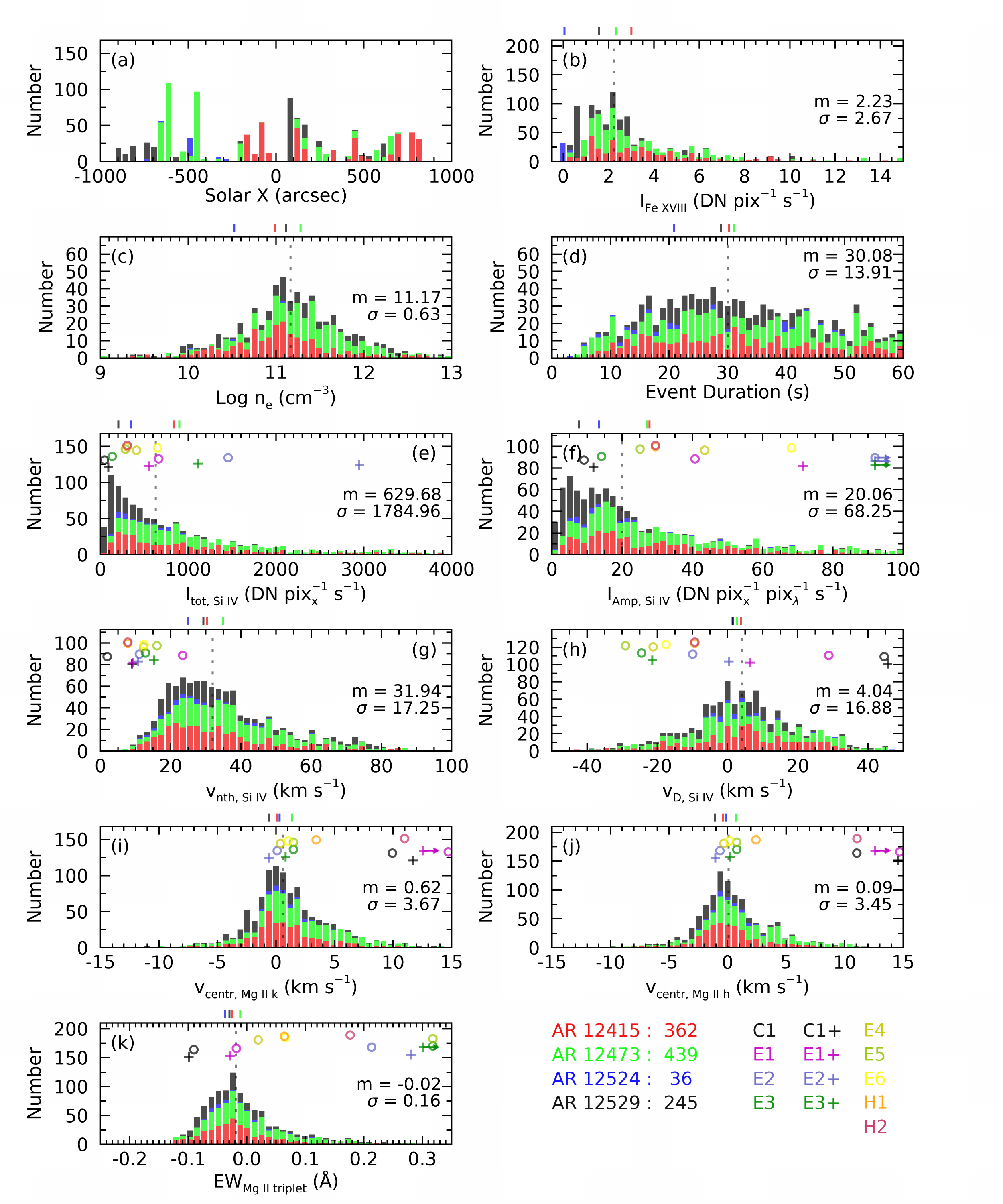}
\caption{Histograms, for the selected footpoint brightenings observed with \iris, of the following parameters: (a) heliocentric $x$ coordinate, (b) mean \fexviii\ intensity for whole \iris\ SJI FOV, (c) electron density from \oiv\ ratio, (d) event duration, (e) total intensity of \siiv\ 1402 \AA\ line, (f) Gaussian amplitude of \siiv\ spectrum, (g) \siiv\ non-thermal velocity, (h) \siiv\ Doppler velocity, (i) \mgii\ k centroids velocity, (j) \mgii\ h centroid velocity, and (k) equivalent width of \mgii\ triplet. Negative Doppler velocity represents blueshift (upward) motion. Each color of histogram indicates a different active region. The median and standard deviation values are written in right corner for each panel. The vertical dashed lines indicate the median value for each parameter. The small colored vertical bars above each panel indicate the median value for the corresponding active region. The circle and cross symbols in the histogram of the \siiv\ and \mgii\ spectral properties represent the values from simulations with arbitrary y offset, and the colors correspond to that of the simulation names shown in the bottom right corner. The cross symbols indicate the results from the simulation model name with +. The right arrow represents values from simulations that are out of the histogram range. The parameters of the simulations are summarized in Table \ref{sim_model}. }
\label{total_hist}
\end{figure*}

\subsection{Distribution of the parameters}
The overall distribution of the obtained parameters is shown in Figure~\ref{total_hist}. The first three parameters (heliocentric $x$ coordinate, \fexviii\ intensity, and the transition region density in Figure~\ref{total_hist}a-c) provide us the background information, and in particular the \fexviii\ emission tracks the hot ($\gtrsim 4$~MK) emission in the active region, and therefore, to some extent, the coronal activity level. The locations of brightenings show that they are widely distributed from heliographic east to west. The \fexviii\ intensity and transition region density have somewhat wide range of values, and depend on the activity level of the active region at that moment. Active region 12524 shows relatively low \fexviii\ intensity and density, while active region 12415 and 12473 show higher values, and this tendency is well matched with their flare productivity in Table \ref{AR_info}. The values for the duration of the events, according to its definition (see section~\ref{sec:analysis}), are almost uniformly distributed from 0 to 60 seconds (Figure \ref{total_hist}d). 

In the histograms of the values of total intensity and amplitude of \siiv\ line (Figure~\ref{total_hist}e, f), it can be seen that the algorithm preferentially found more pixels with smaller values of those parameters. For comparison, the events studied by \citet{Testa2020} show similar or slightly larger ($\gtrsim$ 40 DN pix$_x^{-1}$ pix$_\lambda^{-1}$ s$^{-1}$) \siiv\ amplitudes possibly indicating a bias in their sample toward relatively stronger events because they manually selected the brightening events. The flare productive active regions exhibit higher \siiv\ total intensity and amplitude for their brightenings. The observed ranges for these two parameters are well reproduced by the simulations. Among the simulation models, the thermal conduction (C, C+) and high cutoff energy cases (E3, E5) predict lower \siiv\ intensities. This is not the case for models with intermediate (10~keV) energy cutoff. For former two cases, the thermal conduction and high cutoff energy, the energy is not directly deposited close to the transition region: for the conduction cases, energy will be deposited higher than the transition region, whereas for the higher cutoff energy cases, the energy deposition mostly occurs too deep in the lower atmosphere to significantly affect the \siiv, because the hardness of the electron beam enables the accelerated electrons to reach much deeper regions \citep{Testa2014,Polito2018}.

The non-thermal velocities of the \siiv\ emission line have median $1/e$ width value of about 32 \kms\ with a broad distribution up to $\sim 150$~\kms\ (Figure~\ref{total_hist}g). About half of pixels (53\%) have \siiv\ non-thermal velocity values between 20 and 40 \kms. This result is higher than the previous reported mean value of non-thermal velocity within a whole active region (15 - 20 \kms, \citealt{DePontieu2015}). Here however we selected active locations undergoing brightening, likely explaining the larger non-thermal velocities. We investigated \siiv\ spectral profiles with extremely high non-thermal velocity by visual inspection and found that the \siiv\ line often consists of two or multiple components in those cases. The non-thermal broadening is also higher than we find in the synthetic spectra from simulations. All of the simulation results shows less than 25 \kms, and the synthetic \siiv\ profiles rarely show multiple emission components. We also found that the active region with strong flaring activities shows higher non-thermal velocities, similarly to the findings, discussed above, on  \siiv\ total intensity and amplitude. 
\cite{Testa2020} show a histogram of the increase in \siiv\ non-thermal velocity with respect to its background value outside of the heating event (their Fig.~6), and their values are between 0 and $\sim 40$~\kms, which, considering a background value of the order of 15 - 20~\kms\citep[e.g.,][]{DePontieu2015}, would correspond to a range of $\sim$~15 - 60~\kms\ which matches well the bulk of our observations.

In Figure~\ref{total_hist}h to \ref{total_hist}j, the \siiv\ mean Doppler velocity shows slightly downward motion (redshift of $\simeq 4$ \kms), and the mean values of the \mgii\ k and h centroid Doppler velocities are close to 0 \kms. About 62\% (672/1082), 61\% (657/1082), and 51\% (553/1082) pixels show downward motion in \siiv, \mgii\ k and h lines, respectively. They show almost symmetric distribution with respect to their mean values, and the standard deviation for \siiv\ is about 4.5 times larger than those for the \mgii\ lines, which is reasonable considering the density stratification of the solar atmosphere. These distributions are quite similar for different active regions. The \siiv\ Doppler velocity from synthetic spectra shows a similar range to what is derived from the observations. Only the cases of heating by thermal conduction and low cutoff energy show downward motion in \siiv\ line. This is consistent with the different energy deposition height for different heating mechanisms and heating properties, and analogous to the \mgii\ centroid velocity up to -10 \kms. For the \mgii\ centroids, most models are located near the zero velocity except the above-mentioned two cases (conduction, and 5~keV) and the hybrid models, which have larger downward motions.

The equivalent width values of \mgii\ triplet range from $-$0.1 \AA\ to 0.3 \AA\ with a median value of $-$0.02 \AA\ and a standard deviation of 0.16 \AA (Figure~\ref{total_hist}k). About 36\% (394/1082) of pixels shows positive equivalent width values which correspond to the \mgii\ triplet emission. This finding is also consistent with the results from the small sample of events analyzed by \cite{Testa2020} where they found \mgii\ triplet emission in some locations for 4 out of 10 events. Similar to the \siiv\ parameters, the active region 12473 shows slightly higher equivalent width values than the others. Their distribution is roughly similar to the range of the simulation results. The simulation results tend to have higher equivalent width values with higher cutoff energy and higher flux. The heating by thermal conduction with higher total energy (model C1+) shows lowest equivalent width (EW = $-$0.10), and the heating by accelerated electron with cutoff energy of 15 keV and higher total energy of 10$^{26}$ erg (model E3+) shows highest equivalent width (EW = 0.46). Most of observational results are consistent with the predictions of simulations with lower cutoff energy (E1 and E1+), or intermediate cutoff energy (10~keV) and low energy flux and total energy (E4, E6, and H1).

\subsection{Relation between the parameters}

We analyzed relation between all parameters (See \ref{appendix}). Several relations do not show a strong correlation. Especially, the electron density and the event duration do not exhibit strong correlation with other parameters. However, we found several meaningful relations, and will describe them here below.

\subsubsection{Parameters from the \siiv\ spectra} \label{jpdf:siiv}
\begin{figure*}
\plotone{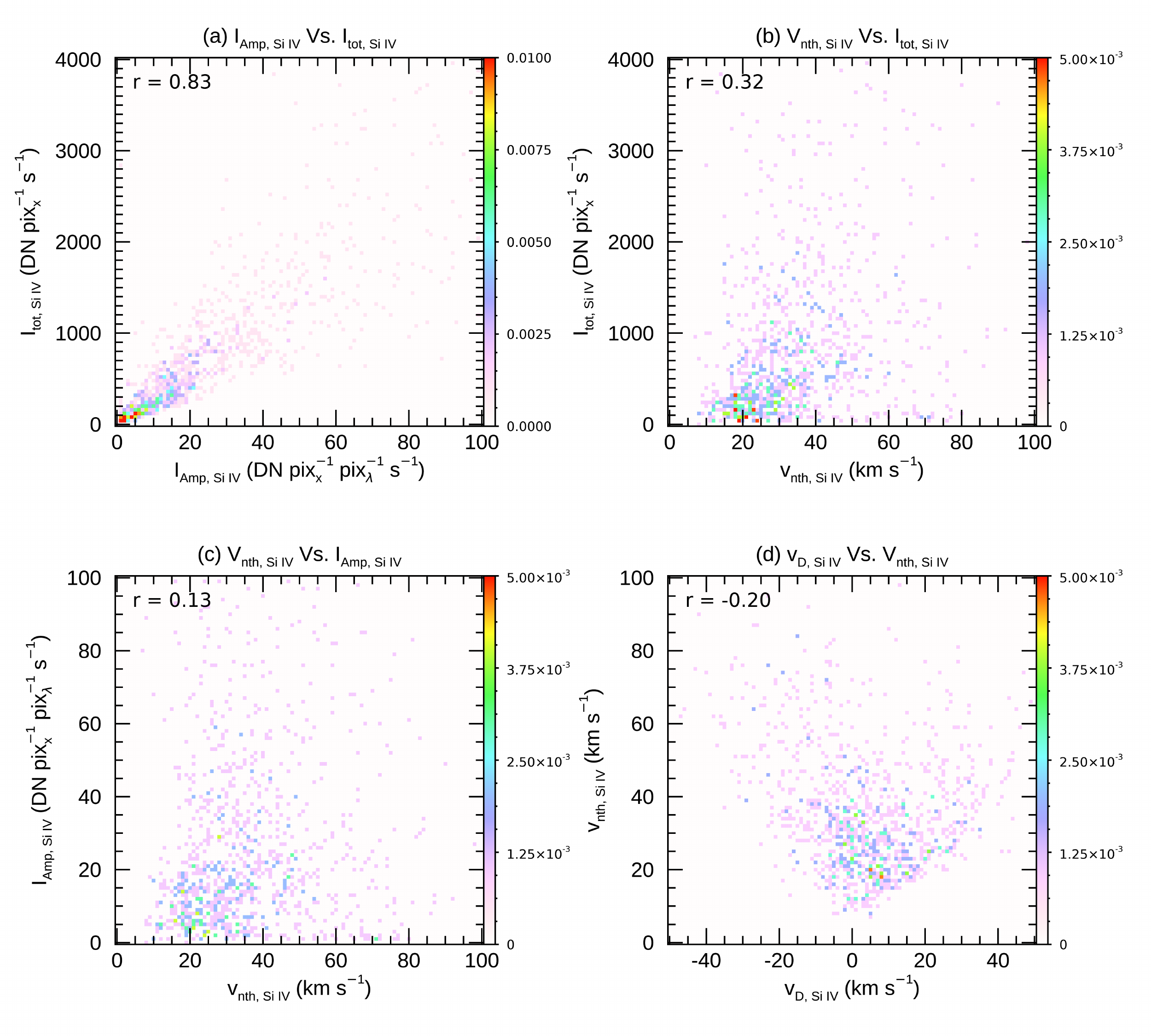}
\caption{Joint probability density function between \siiv\ parameters, (a) amplitude and total intensity, (b) non-thermal velocity and total intensity, (c) non-thermal velocity and amplitude, and (d) Doppler velocity and non-thermal velocity. The colors indicate the probability density which is calculated by dividing the number of data points within each bin by the total number of data points. The Pearson correlation coefficient is shown in top left corner of each panel.}
\label{si_iv_anal}
\end{figure*}

We obtained the total intensity of \siiv\ through integration of the spectrum (see section~\ref{sec:analysis} for details). If we ignore the fitting error, the total intensity of the \siiv\ emission line is approximated by the area of the fitted Gaussian function, and it should be proportional to the Gaussian amplitude and width. In fact, in Fig.~\ref{si_iv_anal}a, the relation between the total intensity and amplitude of the \siiv\ line accordingly shows very strong linear relation with Pearson correlation coefficient of 0.83, as expected if the observed \siiv\ line profiles are well fitted with a Gaussian function Similarly, we found some correlation between the total intensity of the \siiv\ and non-thermal velocity (Fig.~\ref{si_iv_anal}b), as a proxy of the Gaussian width, but the correlation is only moderate and its correlation coefficient is 0.32, since the line width is basically dominated by the thermal component. So the total intensity of \siiv\ line is more dependent on its amplitude than non-thermal velocity. \cite{Testa2016} find a similar moderate correlation between non-thermal width and line intensity for \fexii\ in active region moss.

We also found that the non-thermal velocity is not significantly correlated with the amplitude of \siiv\ line (Fig.~\ref{si_iv_anal}c). This is in contrast with findings of a previous study of the \siiv\ line properties in a whole active region: \citet{DePontieu2015} found a correlation between logarithm of \siiv\ intensity amplitude and non-thermal velocity on large scales. A possible reason for the discrepancy might be the presence of multiple emission components in several \siiv\ spectra. If an emission line has multiple components, its amplitude would relatively decrease while the non-thermal velocity would increase, compared with the corresponding Gaussian with same total intensity. So, two possibilities can coexist for the low amplitude spectrum cases: a simple weak event with small non-thermal velocity, or multiple emission components with large non-thermal velocity. This may be responsible for the observed weak linear correlation between non-thermal velocity and \siiv\ amplitude. Another possibility is that these brightenings have a different behavior from the typical active region locations. It is likely that in the regions, which are undergoing heating, the atmospheric structure differs from that of the average active region. After all, we focus on moss brightenings for which it might be more likely to have multiple emission components (some examples were also observed in \citealt{Testa2020}).

One interesting result is that the joint probability density function between \siiv\ Doppler velocity and non-thermal velocity shows an inverted triangle shape distribution (Fig.~\ref{si_iv_anal}d). If non-thermal velocity is close to zero, the Doppler velocity is also close to zero. For higher non-thermal velocity cases, it is possible to have various Doppler velocity values. It implies that some of non zero Doppler velocities are related to the high non-thermal velocities, i.e., multiple emission components. If so, there is a possibility that the Doppler velocities of moving material may be underestimated. 

\subsubsection{\mgii\ Doppler velocities} \label{jpdf:mgii}
\begin{figure*}
\plotone{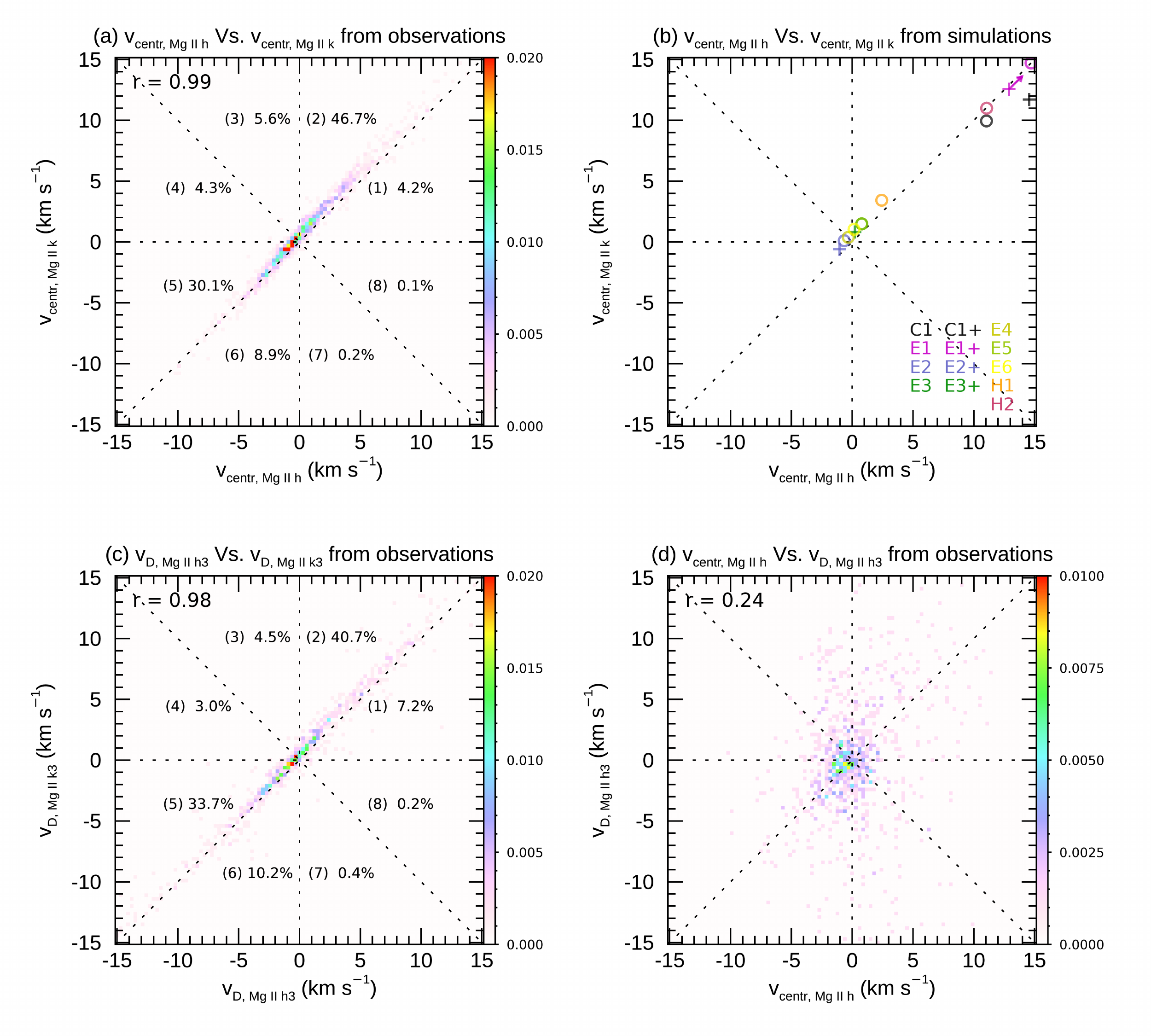}
\caption{(a) Joint probability density function of \mgii\ h \& k centroid velocities from the observations. We divided the parameter space into 8 areas from (1) to (8) and calculate the percentage of the pixels located in each region. The negative velocity represents blueshift and upward velocity. (b) Scatterplot of \mgii\ h \& k centroid velocities from the simulations. The name of simulation models corresponding to each color and symbol are presented at the bottom right corner, as in Figure \ref{total_hist}. (c) Same as (a), but using the \mgii\ h3 and k3 Doppler velocity. (d) Joint probability density function between \mgii\ h centroid velocity and \mgii\ h3 Doppler velocity. The format is same as Figure \ref{si_iv_anal}.
}
\label{mg_ii_anal}
\end{figure*}

In Figure \ref{mg_ii_anal}a, we found that the centroid Doppler velocities in the \mgii\ h\&k lines show extremely strong correlation (r=0.99), as expected considering that their formation heights are very close to each other and overlapping, so it is very likely that dynamic events affect both lines similarly. Additionally, those two \mgii\ line velocities have a good correlation with the \siiv\ Doppler velocity (r=0.64, see \ref{all_density}). This suggests that the heating occurring in the moss brightenings we analyze here often affects similarly the chromosphere and transition region. 
\citet{Leenaarts2013} estimated the formation height of \mgii\ h3 and k3, and of the \mgii\ line peaks, using three-dimentional radiative MHD simulations of quiet Sun. They found that the formation height of these spectral features is generally a little bit higher (about 20-50~km) for the k3 line than for the h3 line. This seems also compatible with the relative standard deviation we derive for their Doppler velocities in Figure \ref{total_hist}i and \ref{total_hist}j. We therefore can conjecture that the k centroid, and k3 Doppler shift, probe slightly higher atmospheric layers compared with the corresponding properties of the h line.

Figure \ref{mg_ii_anal}a shows the relation between \mgii\ h and k Doppler velocities measured by their centroid. It is clear that both velocities have the same sign in most cases. About 51\% and 39.5\% of pixels show both redshifts (area (1) and (2)) or both blueshifts (area (5) and (6)), respectively. The results from the simulations follow a distribution similar to the observational results (Figure \ref{mg_ii_anal}b), although not covering the full observed range, especially with a marked lack of pronounced blueshifts. The models with cutoff energy of 10~keV and 15~keV match well the high density function region (nearby 0 \kms\ for both lines) from the observations. The models related to the thermal conduction (C1, C1+, H1, H2) or low cutoff energy (E1, E1+) show relatively strong redshifts. 

It is noteworthy that most of the observational data is located above the $y=x$ line. It means that the centroid velocity of \mgii\ k line is usually greater than that of \mgii\ h line. To interpret this phenomenon, we consider the upward and downward cases separately. In the downward motion case (area (1) and (2)), for which the energy is conducted from above, considering the direction of progress and formation heights of two lines, the k and h Doppler velocities correspond to the velocity before and after passage of the perturbation through the specific chromospheric layer, respectively. Thus, the fact that most downward pixels show faster k velocity than h velocity (area (2)) implies deceleration. This seems reasonable because the material moves to denser lower region. Similarly, the upward motion case also shows deceleration (area (5)). This may be due to the effects of gravity (ballistic motion) or loss of kinetic energy. 

Interestingly, the correlation between \mgii\ h3 and k3 Doppler velocities shows very similar features (Figure \ref{mg_ii_anal}c). Their Pearson correlation coefficient is 0.98, which is almost the same compared with the coefficient obtained by using the centroid velocities. The h3 and k3 Doppler velocities have the same sign of the h and k centroids, and similarly suggest deceleration in both the upward and downward cases. 
However, we note that the \mgii\ h centroid is not strongly correlated with the \mgii\ h3 velocity (Figure \ref{mg_ii_anal}d), and the same holds true for the \mgii\ k line. We investigated the shapes of the \mgii\ profiles and found that \mgii\ h3 positions are occasionally shifted from the centroid of the h line, i.e., the \mgii\ h spectral line is not symmetric. If h3 Doppler velocity and centroid velocity represent the motion in the upper chromosphere and averaged chromosphere respectively, their weak correlation indicates that the upper choromospheric motions are different from the averaged chromospheric motions, and very complex dynamics characterize the chromosphere when the nanoflares occur.

\subsubsection{\siiv\ non-thermal velocity and $\mu = \cos\theta$} \label{jpdf:nth}
The origin of non-thermal broadening of spectral lines is a long-unsolved mystery. One possibility is that it is caused by unresolved motions. The relation between the non-thermal broadening and the magnetic field direction can provide constraints on the possible mechanisms driving these unresolved motions.
In particular, \alfven\ waves are expected to cause motions perpendicular to the magnetic field direction. We investigated the moss regions in the active regions, i.e., the foot points of hot coronal loops, so we can hypothesize that the magnetic field direction is very likely perpendicular to the solar surface. In our sample we have moss data at various positions on the solar disk with different $\mu = \cos \theta$. Thus, by analyzing the relation between their location from solar disk center and non-thermal velocity, we can obtain some information about the physical origin of the observed non-thermal broadening. 

\begin{figure}
\plotone{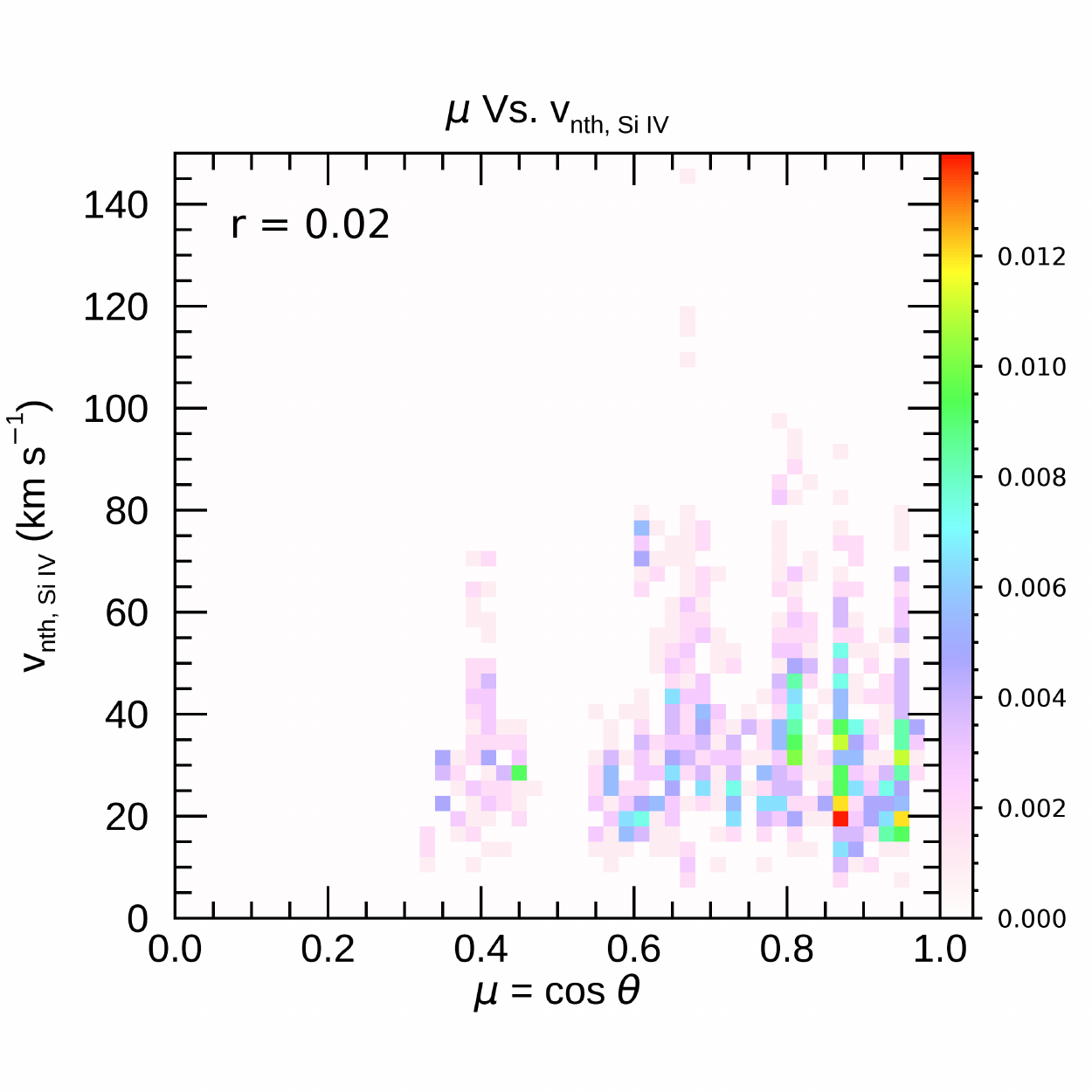}
\caption{The joint probability density function between $\mu = \cos \theta$ and \siiv\ non-thermal velocity. The format is same as Figure \ref{si_iv_anal}.}
\label{mu_nth_anal}
\end{figure}

Our results, shown in Figure~\ref{mu_nth_anal}, demonstrate that the measured non-thermal broadening does not appear to have any clear dependence on the distance from the solar disk center. Most of the \siiv\ non-thermal velocities show values between 15 and 40 \kms\ regardless of the $\mu$ values. The Pearson correlation coefficient is 0.02, which means that two parameters do not have a linear correlation. 
These findings suggest that the non-thermal broadening is not related to the direction of magnetic fields, hinting that \alfven\ waves do not provide significant contribution to the heating in these events.

\subsubsection{The \mgii\ triplet emission} \label{jpdf:mgii_trip}
\begin{figure*}
\plotone{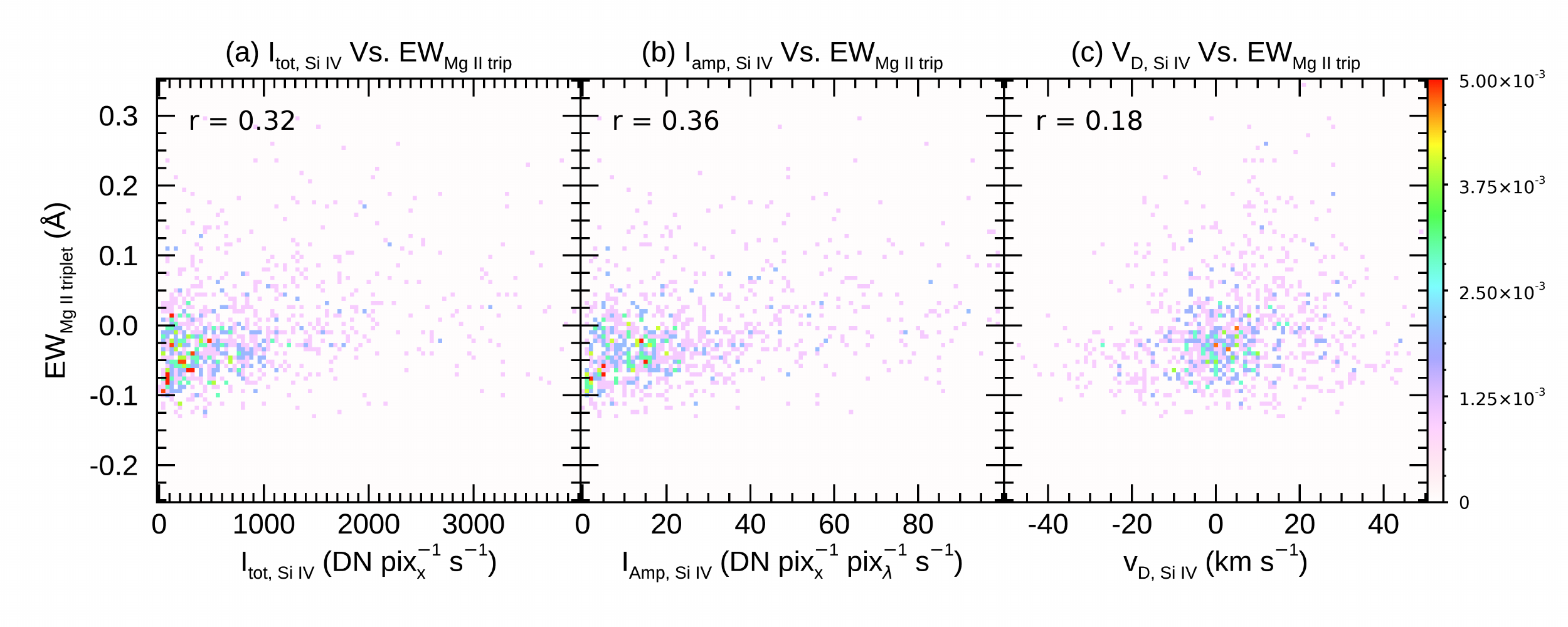}
\caption{Joint probability density function between the equivalent width of \mgii\ triplet and \siiv\ parameters, (a) total intensity, (b) Gaussian amplitude, and (c) Doppler velocity. The format is same as Figure \ref{si_iv_anal}.}
\label{mg_trip_anal}
\end{figure*}


The \mgii\ triplet equivalent width is correlated with the total intensity and the Gaussian amplitude of the \siiv\ line (Figure \ref{mg_trip_anal}a and \ref{mg_trip_anal}b). Those two parameters from the \siiv\ have a strong correlation with each other (in Section \ref{jpdf:siiv}), therefore it is not surprising that if a correlation with the \mgii\ triplet equivalent width is found, it is found for both parameters with similar correlation coefficients. According to the theoretical calculations and numerical simulations, the \mgii\ triplet is formed in the low chromosphere, and it is in emission in presence of a steep temperature increase in the lower chromosphere \citep{Pereira2015}, while the \siiv\ is formed in the lower transition region \citep[e.g.,][]{Dudik2014}. The RADYN simulations also predict a weak correlation between \mgii\ triplet equivalent width and line intensity as observed: non-thermal electrons of sufficient energy will penetrate deeper in the atmosphere and deposit their energy in the chromosphere simultaneously causing \mgii\ triplet emission and large \siiv\ emission, while heating in the corona and subsequent conduction (or non-thermal electron with low energy $\lesssim 5$~keV) produces weaker \siiv\ emission, and no significant heating in the lower chromosphere, and therefore negative \mgii\ equivalent width. However, from the models we would also expect a negative correlation of the \mgii\ triplet emission with the \siiv\ Doppler shift, which is not evident in the observations (Figure \ref{mg_trip_anal}c).

\subsubsection{\fexviii\ intensity} \label{jpdf:fe_xviii}
\begin{figure*}
\plotone{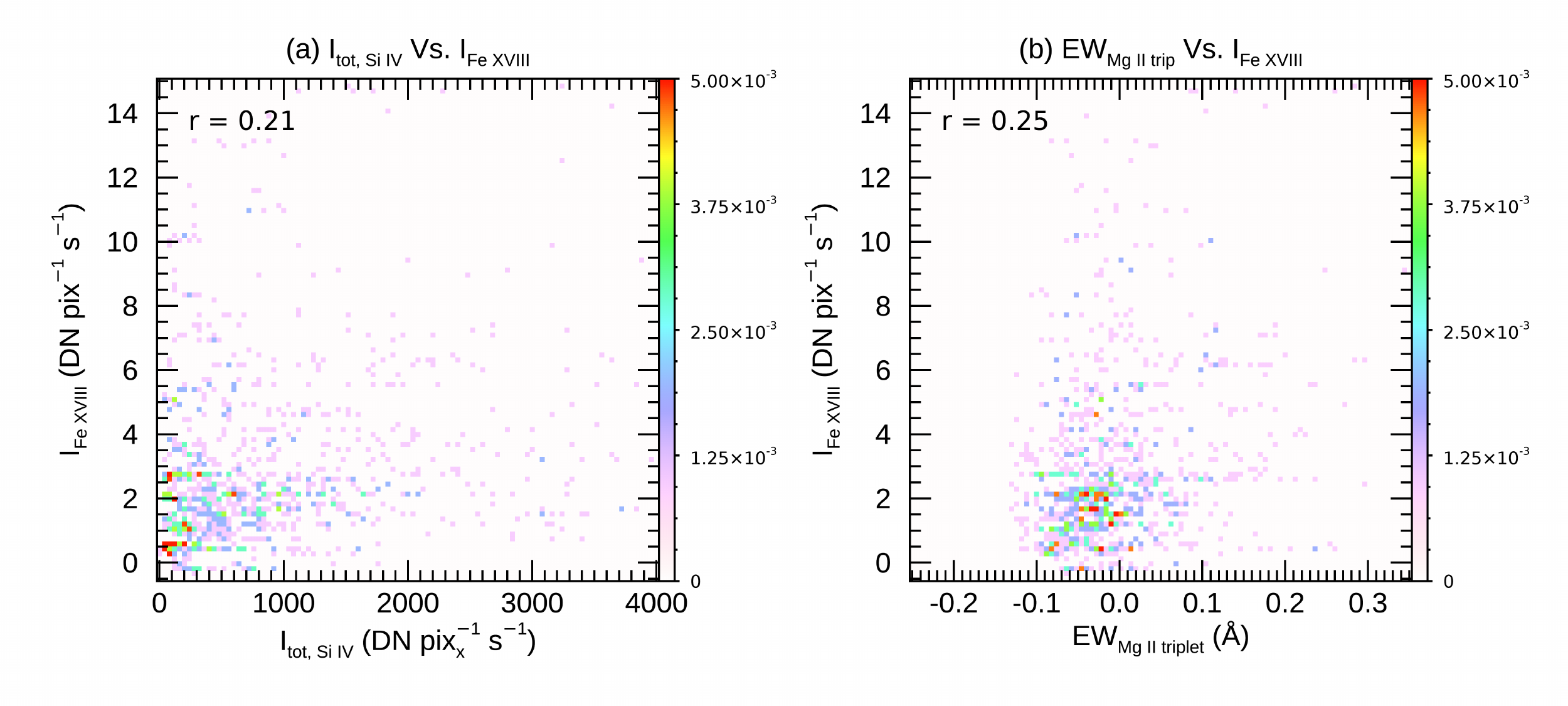}
\caption{Joint probability density function between the \fexviii\ intensity and (a) \siiv\ total intensity and (b) equivalent width of \mgii\ triplet. The format is same as Figure \ref{si_iv_anal}.
}
\label{fe_xviii_anal}
\end{figure*}

The \fexviii\ emission is a representative parameter reflecting the amount of coronal plasma hotter than $\sim 4$~MK, and clearly correlates with the activity level of the active region, as can be deduced from Figure~\ref{total_hist}b. However, the \fexviii\ emission value we derive is spatially averaged (in the \iris\ FOV, see sect.~\ref{sec:analysis}), at the time of the moss brightening and therefore can provide some information on the possible effect of the instantaneous coronal environment on the properties of the heating. 

We find that the \fexviii\ emission shows some correlation, although weak, with the total intensity of \siiv\ and with the equivalent width of the \mgii\ triplet (Figure \ref{fe_xviii_anal}). As discussed in the previous section, our RADYN simulations suggest that these two parameters have some dependence on the parameters of the heating, and in particular that they are enhanced when accelerated electrons of sufficiently high energy (i.e., low-energy cutoff $\gtrsim 10$~keV) are present, because these non-thermal particles efficiently heat the lower atmosphere. 
Therefore, we can speculate that active regions producing more hot plasma, might more easily accelerate electrons to high energies.

\section{Conclusion and Discussion}
The transition region response to coronal heating events provides very powerful diagnostics of the heating properties. Here we have carried out a statistical study of \iris\ spectral observations of moss brightenings caused by small-scale heating events (nanoflares). We have further developed and algorithm to detect the moss brightenings automatically from the SDO/AIA and \iris\ data \citep{Graham2019}. We have found more than 1000 moss brightenings, and analyzed the \iris\ \siiv\ and \mgii\ emission, extracting their spectral parameters, and analyzing their distribution and the relations among different parameters. 

The automatic detection algorithm allowed us to detect a broader range of events, compared with the manually selected sample we analyzed in previous work \citep{Testa2014,Testa2020}. 
Overall, we find that the parameters derived from our much large sample of events are mostly in agreement with the previous smaller studies. 
In particular, we note that the \siiv\ Doppler velocity has a similar, roughly symmetric, distribution between approximately $-40$ and $40$~\kms, and that the \mgii\ triplet emission (here identified by positive values for the equivalent width of the line) is found for a non-negligible fraction of the sample ($\gtrsim 30$~\%). 
Some observed properties also show some differences. For instance the non-thermal broadening distribution of our sample covers a much broader range of values than previous work. We note that in \citet{Testa2020}, rather than analyzing the absolute values of non-thermal broadening, we had derived the relative increases compared to the quiescent spectra outside of the brightenings. By taking that into account, the values found in the smaller sample of \citet{Testa2020} would correspond to a range of roughly 15 to 60~\kms, not too dissimilar to the bulk of our distribution, although their distribution peaked at the lower end, while we find a mean value of $\sim 32$~\kms.   
Also, we detected a large number of weaker events, as indicated by the histogram showing the distribution of \siiv\ total intensity (Figure \ref{total_hist}e) and amplitude (Figure \ref{total_hist}f). \citet{Testa2020,Testa2014} reported about a dozen of transient moss brightenings. For several of their events the maximum amplitude of the \siiv\ was $\gtrsim$ 40 DN pix$_x^{-1}$ pix$_\lambda^{-1}$ s$^{-1}$, slightly larger than our results with mean value of about 30 DN pix$_x^{-1}$ pix$_\lambda^{-1}$ s$^{-1}$. This preference for weaker events from the automatic detection algorithm is presumably due to the fact that smaller events occur more frequently. It may indicate that the brightenings we found are scaled-down version of large flares, and their occurrence rate perhaps follow a power law \citep[e.g.,][]{Aschwanden2000}. We briefly tested that the \siiv\ total intensity roughly follows power law distribution with power of -0.71, however, further studies are required to obtain the total energy of nanoflare from the \siiv\ spectral data.

We examined the dependence of the moss brightenings on the coronal environment. In Figure \ref{total_hist}b, the flare productive active regions show relatively higher \fexviii\ intensity which indicates the existence of a larger amount of hot plasma in the corona. The \fexviii\ emission shows a moderate positive correlation with the total intensity of \siiv\ line, and with the equivalent width of the \mgii\ triplet line (Figure \ref{fe_xviii_anal}), which can be used as a rough indicator of the cutoff energy of the electron beam, according to the numerical simulations (Figure~\ref{total_hist}k, and  \citealt{Testa2020}). In addition, we found that the moss brightenings occurred more frequently in the flare productive active region. Thus, we conjecture that in the vigorous active regions with hot plasma, higher energy electron beams can be generated more frequently. 

As noted above, compared with our previous work based on small samples of these heating events, here we derived more parameters, and the size of our sample allowed for statistical analysis of the relations between the parameters. 
For the \siiv\ line (Figure \ref{si_iv_anal}), we found that the total intensity is well correlated with the Gaussian amplitude of the line, as expected. In our sample we found no significant correlation between non-thermal velocity and \siiv\ Gaussian amplitude, in contrast with results of previous work on an entire active region \citep{DePontieu2015}. As we discussed above (sec.~\ref{jpdf:siiv}), this might be explained by the significant presence of multiple emission components or because the atmospheric structure disturbed by accelerated particles differs from that of the typical active regions.
Another very interesting correlation we found is between the chromospheric (\mgii\ h\&k) and the transition region Doppler velocities, which show significant positive correlation (Figure \ref{mg_ii_anal}).
This finding points to the fact that the heating event might impact the different layer of the atmosphere similarly. In particular, the \mgii\ h and k line have a very strong correlation of velocities, either measured by the line centroid or the position of the h3/k3 line. In both cases the velocity of the k line (formed at slightly larger heights than h) is slightly larger than for the h line, for both redshifted and blueshifted cases, suggesting a deceleration in both cases. However, it is  interesting to note that the velocities from the \mgii\ centroid or the h3/k3 position only show a modest correlation (r=0.27) indicating that the \mgii\ h\&k lines are mostly not symmetric, and, in turn, this suggests that these heating events are characterized by strong flows and gradients in the chromosphere.

To investigate the nature of the non-thermal velocity, we analyzed its relation with the distance from the solar disk center, i.e., with the inclination with respect to the line-of-sight. We found that the non-thermal velocity does not have a significant correlation with the distance from the solar disk center (Figure \ref{mu_nth_anal}). Previous studies reported that center-to-limb variation of non-thermal velocity or line width is negligible \citep[e.g.,][]{Chae1998, Ghosh2021}. Our work confirms this lack of correlation is also valid in moss brightening regions. This could occur because the turbulent motions are isotropic, or there may be different processes that act along the field and perpendicular to the field and that are of equal magnitude \citep{Tian2014, DePontieu2014b, DePontieu2015}. 

Another finding about the \siiv\ non-thermal velocity is that the observed values are higher than numerical simulations, and than previous reports. The median value of our measurements is 32 \kms, and the distribution has a long tail to values that are higher than 60 \kms. In contrast, the simulation results show less than 25 \kms\ (Figure \ref{total_hist}g). This may be attributed to the fact that the simulations do not contain turbulent motions and assumed a single loop, while several thin adjacent magnetic strands, heated around the same time, are likely observed within a single \iris\ pixel. Clearly such superposition within one pixel can lead to an increase in non-thermal broadening, given the different velocities along independently heated loops.  
In \citet{DePontieu2015}, they measured the non-thermal velocities of \siiv\ line in an active region, and reported that their distribution is close to a normal distribution with peak value of about 18 \kms. Our selected events are for moss regions, heated impulsively by conduction or accelerated electron beams, which might contribute to the high non-thermal broadening tail observed in active regions \citep{DePontieu2015,Testa2016}. In some of the \siiv\ spectra in our sample, the very high non-thermal velocities are due to the presence of multiple components.

Thus, it is clear that the moss brightenings are usually characterized by violent dynamics. This is evidenced not only by the non-thermal velocity of the \siiv\ line, but also by  the asymmetric profiles of \mgii\ spectral lines. An asymmetric profile implies that the h3 or k3 position is shifted with respect to the line centroid, so the upper chromospheric motion is not consistent with the average chromospheric motion. Moreover, the difference of h2v and h2r intensities indicates strong upward or downward motion \citep{Leenaarts2013, Carlsson1997b}.

By comparison with the numerical simulations, our results help to understand the nanoflare mechanism and provide constraint on the physical parameters. Our numerical simulations with various model parameters reproduce fairly well the \siiv\ and \mgii\ spectral properties when nanoflares occur. Most of the parameters have similar values to the observations. In agreement with our previous work \citep{Testa2014,Polito2018,Testa2020}, the observed \siiv\ blueshifted profiles, as well as the \mgii\ triplet emission are only reproduced by simulations including heating by beams of accelerated electrons. 
However, several parameters deviate from the observational data, pointing to interesting issues that need to be investigated in more details, and to the need for additional models that can more consistently explain the observations. Some examples include the relatively small non-thermal velocity, and the lack of significantly large negative value of the \mgii\ centroid velocities in the simulations (Figure \ref{total_hist}g, \ref{total_hist}i and \ref{total_hist}j). As we speculated earlier in the paper, the former might be due to the fact that the single loop models might not be an adequate comparison for the observations in each \iris\ pixel where many strands, possibly with significantly different initial conditions and heating properties, might overlap. The latter might be due to shortcomings of the model in reproducing realistic chromospheric conditions as discussed at length in previous literature \citep[e.g., see discussions in][]{Polito2018}.
Another puzzling discrepancy between simulations and observations is the relation between the \siiv\ Doppler velocity and the equivalent width of the \mgii\ triplet. The  numerical simulations characterized by blueshifts also have stronger \mgii\ triplet emission, whereas the two parameters appear to have a very weak but \emph{positive} correlation in the observational data (Figure \ref{mg_trip_anal}c). It is possible that our definition of the equivalent width does not adequately capture \mgii\ emission in the observations, but this discrepancy will be thoroughly investigated in future work.

In summary this work has presented a thorough analysis of the observational signatures of the lower atmospheric (chromosphere and transition region) response to coronal small heating events. The comparison with RADYN simulations of nanoflare-heated loops has highlighted the overall agreement between the predictions and the observations for individual, single observables (e.g., intensity). However, we have also found some discrepancies, and there are no specific models that explain all observed parameters simultaneously. These limitations shed light on the shortcoming of the models and emphasize the need for more diverse and realistic numerical simulations for a more comprehensive reproduction of the observed signatures and interpretation of the data.


\vspace{1cm}
\begin{acknowledgements}
We gratefully acknowledge support by the NASA contract NNG09FA40C (IRIS). 
This research has made use of NASA's Astrophysics Data System and of the SolarSoft package for IDL.
\iris\ is a NASA small explorer mission developed and operated by LMSAL with mission operations executed at NASA Ames Research Center and major contributions to downlink communications funded by ESA and the Norwegian Space Centre. Resources supporting this work were provided by the NASA High-End Computing (HEC) Program through the NASA Advanced Supercomputing (NAS) Division at Ames Research Center. CHIANTI is a collaborative project involving George Mason University, the University of Michigan (USA), University of Cambridge (UK) and NASA Goddard Space Flight Center (USA).
\end{acknowledgements}

\appendix 
\restartappendixnumbering 
\section{Joint probability density function for all parameters} \label{appendix}
\begin{figure*}
\includegraphics[width=1\textwidth]{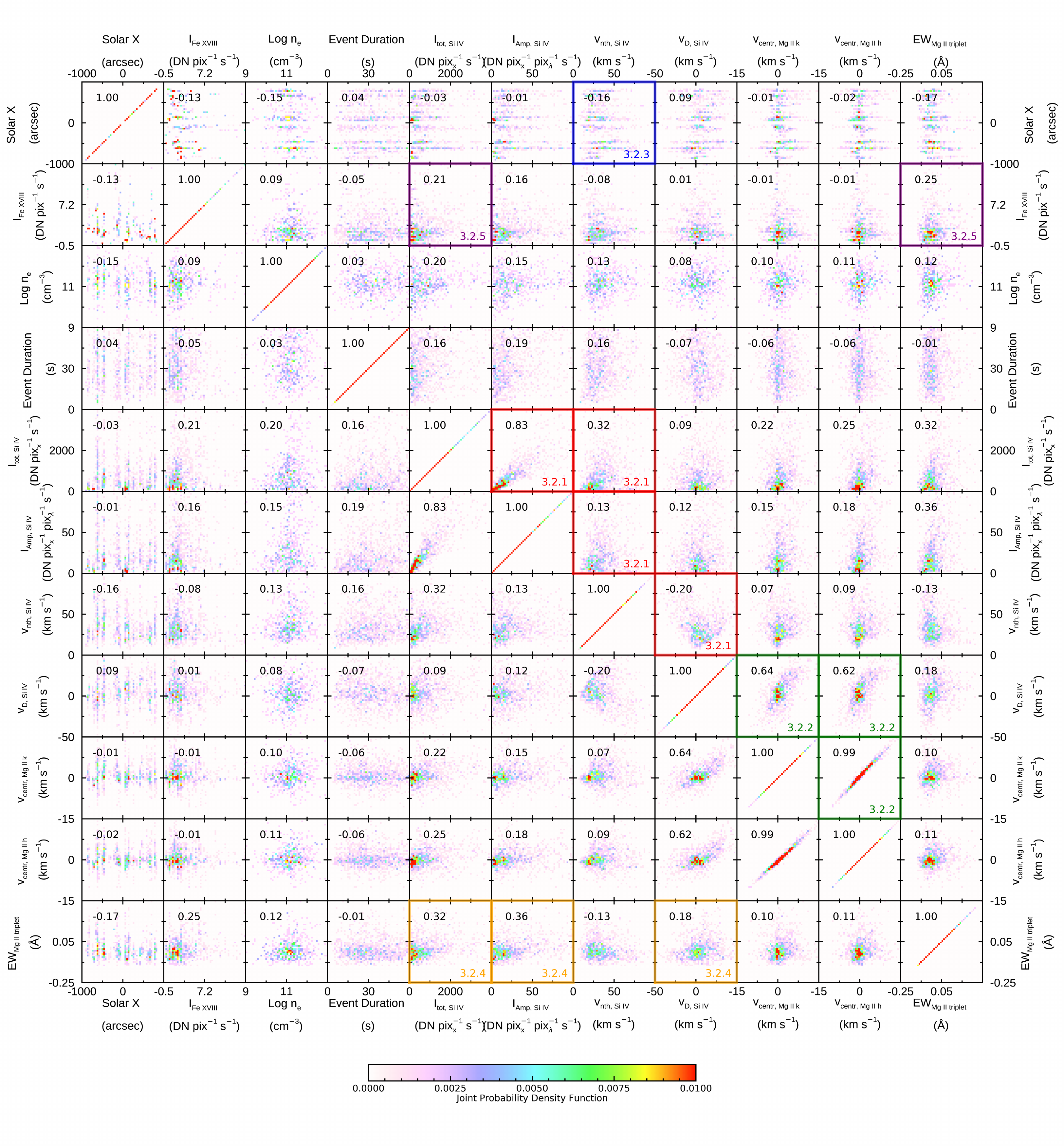}
\caption{Joint probability density function for all parameters. The image is diagonally symmetric, but we show the entire matrix of combinations for the reader's convenience. The Pearson correlation coefficient between two parameters is shown in the top left corner of each plot. The colored boxes with red, green, blue, orange, purple are related to section~\ref{jpdf:siiv} to \ref{jpdf:fe_xviii}, respectively.}
\label{all_density}
\end{figure*}

\bibliography{PASPsample631}{}

\begin{thebibliography}{}
\expandafter\ifx\csname natexlab\endcsname\relax\def\natexlab#1{#1}\fi
\providecommand{\url}[1]{\href{#1}{#1}}
\providecommand{\dodoi}[1]{doi:~\href{http://doi.org/#1}{\nolinkurl{#1}}}
\providecommand{\doeprint}[1]{\href{http://ascl.net/#1}{\nolinkurl{http://ascl.net/#1}}}
\providecommand{\doarXiv}[1]{\href{https://arxiv.org/abs/#1}{\nolinkurl{https://arxiv.org/abs/#1}}}

\bibitem[{{Allred} {et~al.}(2015){Allred}, {Kowalski}, \&
  {Carlsson}}]{Allred2015}
{Allred}, J.~C., {Kowalski}, A.~F., \& {Carlsson}, M. 2015, \apj, 809, 104,
  \dodoi{10.1088/0004-637X/809/1/104}

\bibitem[{{Antiochos} {et~al.}(2003){Antiochos}, {Karpen}, {DeLuca}, {Golub},
  \& {Hamilton}}]{Antiochos2003}
{Antiochos}, S.~K., {Karpen}, J.~T., {DeLuca}, E.~E., {Golub}, L., \&
  {Hamilton}, P. 2003, \apj, 590, 547

\bibitem[{{Aschwanden} {et~al.}(2000){Aschwanden}, {Tarbell}, {Nightingale},
  {Schrijver}, {Title}, {Kankelborg}, {Martens}, \& {Warren}}]{Aschwanden2000}
{Aschwanden}, M.~J., {Tarbell}, T.~D., {Nightingale}, R.~W., {et~al.} 2000,
  \apj, 535, 1047, \dodoi{10.1086/308867}

\bibitem[{{Bakke} {et~al.}(2022){Bakke}, {Carlsson}, {Rouppe van der Voort},
  {Gudiksen}, {Polito}, {Testa}, \& {De Pontieu}}]{Bakke2022}
{Bakke}, H., {Carlsson}, M., {Rouppe van der Voort}, L., {et~al.} 2022, \aap,
  659, A186, \dodoi{10.1051/0004-6361/202142842}

\bibitem[{{Carlsson} \& {Stein}(1992)}]{Carlsson1992}
{Carlsson}, M., \& {Stein}, R.~F. 1992, \apjl, 397, L59, \dodoi{10.1086/186544}

\bibitem[{{Carlsson} \& {Stein}(1995)}]{Carlsson1995}
---. 1995, \apjl, 440, L29, \dodoi{10.1086/187753}

\bibitem[{{Carlsson} \& {Stein}(1997{\natexlab{a}})}]{Carlsson1997a}
---. 1997{\natexlab{a}}, {Chromospheric Dynamics - What Can Be Learnt from
  Numerical Simulations}, ed. G.~M. {Simnett}, C.~E. {Alissandrakis}, \&
  L.~{Vlahos}, Vol. 489, 159, \dodoi{10.1007/BFb0105675}

\bibitem[{{Carlsson} \& {Stein}(1997{\natexlab{b}})}]{Carlsson1997b}
---. 1997{\natexlab{b}}, \apj, 481, 500, \dodoi{10.1086/304043}

\bibitem[{Chae {et~al.}(1998)Chae, Sch{\"u}hle, \& Lemaire}]{Chae1998}
Chae, J., Sch{\"u}hle, U., \& Lemaire, P. 1998, The Astrophysical Journal, 505,
  957

\bibitem[{{Cheung} {et~al.}(2015){Cheung}, {Boerner}, {Schrijver}, {Testa},
  {Chen}, {Peter}, \& {Malanushenko}}]{Cheung2015}
{Cheung}, M. C.~M., {Boerner}, P., {Schrijver}, C.~J., {et~al.} 2015, \apj,
  807, 143, \dodoi{10.1088/0004-637X/807/2/143}

\bibitem[{{Cooper} {et~al.}(2021){Cooper}, {Hannah}, {Grefenstette},
  {Glesener}, {Krucker}, {Hudson}, {White}, {Smith}, \& {Duncan}}]{Cooper2021}
{Cooper}, K., {Hannah}, I.~G., {Grefenstette}, B.~W., {et~al.} 2021, \mnras,
  507, 3936, \dodoi{10.1093/mnras/stab2283}

\bibitem[{{Curdt} {et~al.}(2004){Curdt}, {Landi}, \& {Feldman}}]{Curdt2004}
{Curdt}, W., {Landi}, E., \& {Feldman}, U. 2004, \aap, 427, 1045,
  \dodoi{10.1051/0004-6361:20041278}

\bibitem[{{De Pontieu} {et~al.}(1999){De Pontieu}, {Berger}, {Schrijver}, \&
  {Title}}]{DePontieu1999}
{De Pontieu}, B., {Berger}, T.~E., {Schrijver}, C.~J., \& {Title}, A.~M. 1999,
  \solphys, 190, 419, \dodoi{10.1023/A:1005220606223}

\bibitem[{{De Pontieu} {et~al.}(2015){De Pontieu}, {McIntosh},
  {Martinez-Sykora}, {Peter}, \& {Pereira}}]{DePontieu2015}
{De Pontieu}, B., {McIntosh}, S., {Martinez-Sykora}, J., {Peter}, H., \&
  {Pereira}, T.~M.~D. 2015, \apjl, 799, L12,
  \dodoi{10.1088/2041-8205/799/1/L12}

\bibitem[{{De Pontieu} {et~al.}(2014){De Pontieu}, {Title}, {Lemen}, {Kushner},
  {Akin}, {Allard}, {Berger}, {Boerner}, {Cheung}, {Chou}, {Drake}, {Duncan},
  {Freeland}, {Heyman}, {Hoffman}, {Hurlburt}, {Lindgren}, {Mathur}, {Rehse},
  {Sabolish}, {Seguin}, {Schrijver}, {Tarbell}, {W{\"u}lser}, {Wolfson},
  {Yanari}, {Mudge}, {Nguyen-Phuc}, {Timmons}, {van Bezooijen}, {Weingrod},
  {Brookner}, {Butcher}, {Dougherty}, {Eder}, {Knagenhjelm}, {Larsen},
  {Mansir}, {Phan}, {Boyle}, {Cheimets}, {DeLuca}, {Golub}, {Gates}, {Hertz},
  {McKillop}, {Park}, {Perry}, {Podgorski}, {Reeves}, {Saar}, {Testa}, {Tian},
  {Weber}, {Dunn}, {Eccles}, {Jaeggli}, {Kankelborg}, {Mashburn}, {Pust},
  {Springer}, {Carvalho}, {Kleint}, {Marmie}, {Mazmanian}, {Pereira}, {Sawyer},
  {Strong}, {Worden}, {Carlsson}, {Hansteen}, {Leenaarts}, {Wiesmann},
  {Aloise}, {Chu}, {Bush}, {Scherrer}, {Brekke}, {Martinez-Sykora}, {Lites},
  {McIntosh}, {Uitenbroek}, {Okamoto}, {Gummin}, {Auker}, {Jerram}, {Pool}, \&
  {Waltham}}]{DePontieu2014}
{De Pontieu}, B., {Title}, A.~M., {Lemen}, J.~R., {et~al.} 2014, \solphys, 289,
  2733, \dodoi{10.1007/s11207-014-0485-y}

\bibitem[{De~Pontieu {et~al.}(2014)De~Pontieu, Rouppe van~der Voort, McIntosh,
  Pereira, Carlsson, Hansteen, Skogsrud, Lemen, Title, Boerner,
  {et~al.}}]{DePontieu2014b}
De~Pontieu, B., Rouppe van~der Voort, L., McIntosh, S.~W., {et~al.} 2014,
  Science, 346, 1255732

\bibitem[{{Del Zanna}(2013)}]{DelZanna2013}
{Del Zanna}, G. 2013, \aap, 558, A73, \dodoi{10.1051/0004-6361/201321653}

\bibitem[{{Dud{\'\i}k} {et~al.}(2014){Dud{\'\i}k}, {Del Zanna},
  {Dzif{\v{c}}{\'a}kov{\'a}}, {Mason}, \& {Golub}}]{Dudik2014}
{Dud{\'\i}k}, J., {Del Zanna}, G., {Dzif{\v{c}}{\'a}kov{\'a}}, E., {Mason},
  H.~E., \& {Golub}, L. 2014, \apjl, 780, L12,
  \dodoi{10.1088/2041-8205/780/1/L12}

\bibitem[{{Ghosh} {et~al.}(2021){Ghosh}, {Tripathi}, \& {Klimchuk}}]{Ghosh2021}
{Ghosh}, A., {Tripathi}, D., \& {Klimchuk}, J.~A. 2021, \apj, 913, 151,
  \dodoi{10.3847/1538-4357/abf244}

\bibitem[{{Glesener} {et~al.}(2020){Glesener}, {Krucker}, {Duncan}, {Hannah},
  {Grefenstette}, {Chen}, {Smith}, {White}, \& {Hudson}}]{Glesener2020}
{Glesener}, L., {Krucker}, S., {Duncan}, J., {et~al.} 2020, \apjl, 891, L34,
  \dodoi{10.3847/2041-8213/ab7341}

\bibitem[{{Graham} {et~al.}(2019){Graham}, {De Pontieu}, \&
  {Testa}}]{Graham2019}
{Graham}, D.~R., {De Pontieu}, B., \& {Testa}, P. 2019, \apjl, 880, L12,
  \dodoi{10.3847/2041-8213/ab2f91}

\bibitem[{{Hale} {et~al.}(1919){Hale}, {Ellerman}, {Nicholson}, \&
  {Joy}}]{Hale1919}
{Hale}, G.~E., {Ellerman}, F., {Nicholson}, S.~B., \& {Joy}, A.~H. 1919, \apj,
  49, 153, \dodoi{10.1086/142452}

\bibitem[{{Hannah} {et~al.}(2008){Hannah}, {Christe}, {Krucker}, {Hurford},
  {Hudson}, \& {Lin}}]{Hannah2008}
{Hannah}, I.~G., {Christe}, S., {Krucker}, S., {et~al.} 2008, \apj, 677, 704,
  \dodoi{10.1086/529012}

\bibitem[{{Holman} {et~al.}(2011){Holman}, {Aschwanden}, {Aurass}, {Battaglia},
  {Grigis}, {Kontar}, {Liu}, {Saint-Hilaire}, \& {Zharkova}}]{Holman2011}
{Holman}, G.~D., {Aschwanden}, M.~J., {Aurass}, H., {et~al.} 2011, \ssr, 159,
  107, \dodoi{10.1007/s11214-010-9680-9}

\bibitem[{{Klimchuk}(2006)}]{Klimchuk2006}
{Klimchuk}, J.~A. 2006, \solphys, 234, 41, \dodoi{10.1007/s11207-006-0055-z}

\bibitem[{{K{\"u}nzel}(1965)}]{Kunzel1965}
{K{\"u}nzel}, H. 1965, Astronomische Nachrichten, 288, 177

\bibitem[{{Leenaarts} {et~al.}(2013){Leenaarts}, {Pereira}, {Carlsson},
  {Uitenbroek}, \& {De Pontieu}}]{Leenaarts2013}
{Leenaarts}, J., {Pereira}, T.~M.~D., {Carlsson}, M., {Uitenbroek}, H., \& {De
  Pontieu}, B. 2013, \apj, 772, 90, \dodoi{10.1088/0004-637X/772/2/90}

\bibitem[{{Lemen} {et~al.}(2012){Lemen}, {Title}, {Akin}, {Boerner}, {Chou},
  {Drake}, {Duncan}, {Edwards}, {Friedlaender}, {Heyman}, {Hurlburt}, {Katz},
  {Kushner}, {Levay}, {Lindgren}, {Mathur}, {McFeaters}, {Mitchell}, {Rehse},
  {Schrijver}, {Springer}, {Stern}, {Tarbell}, {Wuelser}, {Wolfson}, {Yanari},
  {Bookbinder}, {Cheimets}, {Caldwell}, {Deluca}, {Gates}, {Golub}, {Park},
  {Podgorski}, {Bush}, {Scherrer}, {Gummin}, {Smith}, {Auker}, {Jerram},
  {Pool}, {Soufli}, {Windt}, {Beardsley}, {Clapp}, {Lang}, \&
  {Waltham}}]{Lemen2012}
{Lemen}, J.~R., {Title}, A.~M., {Akin}, D.~J., {et~al.} 2012, \solphys, 275,
  17, \dodoi{10.1007/s11207-011-9776-8}

\bibitem[{{Parker}(1988)}]{Parker1988}
{Parker}, E.~N. 1988, \apj, 330, 474, \dodoi{10.1086/166485}

\bibitem[{{Pereira} {et~al.}(2015){Pereira}, {Carlsson}, {De Pontieu}, \&
  {Hansteen}}]{Pereira2015}
{Pereira}, T. M.~D., {Carlsson}, M., {De Pontieu}, B., \& {Hansteen}, V. 2015,
  \apj, 806, 14, \dodoi{10.1088/0004-637X/806/1/14}

\bibitem[{{Pereira} \& {Uitenbroek}(2015)}]{PereiraUitenbroek2015}
{Pereira}, T. M.~D., \& {Uitenbroek}, H. 2015, \aap, 574, A3,
  \dodoi{10.1051/0004-6361/201424785}

\bibitem[{{Pesnell} {et~al.}(2012){Pesnell}, {Thompson}, \&
  {Chamberlin}}]{Pesnell2012}
{Pesnell}, W.~D., {Thompson}, B.~J., \& {Chamberlin}, P.~C. 2012, \solphys,
  275, 3, \dodoi{10.1007/s11207-011-9841-3}

\bibitem[{{Polito} {et~al.}(2018){Polito}, {Testa}, {Allred}, {De Pontieu},
  {Carlsson}, {Pereira}, {Go{\v{s}}i{\'c}}, \& {Reale}}]{Polito2018}
{Polito}, V., {Testa}, P., {Allred}, J., {et~al.} 2018, \apj, 856, 178,
  \dodoi{10.3847/1538-4357/aab49e}

\bibitem[{{Schmit} {et~al.}(2015){Schmit}, {Bryans}, {De Pontieu}, {McIntosh},
  {Leenaarts}, \& {Carlsson}}]{Schmit2015}
{Schmit}, D., {Bryans}, P., {De Pontieu}, B., {et~al.} 2015, \apj, 811, 127,
  \dodoi{10.1088/0004-637X/811/2/127}

\bibitem[{{Testa} {et~al.}(2016){Testa}, {De Pontieu}, \&
  {Hansteen}}]{Testa2016}
{Testa}, P., {De Pontieu}, B., \& {Hansteen}, V. 2016, \apj, 827, 99,
  \dodoi{10.3847/0004-637X/827/2/99}

\bibitem[{{Testa} {et~al.}(2020){Testa}, {Polito}, \& {Pontieu}}]{Testa2020}
{Testa}, P., {Polito}, V., \& {Pontieu}, B.~D. 2020, \apj, 889, 124,
  \dodoi{10.3847/1538-4357/ab63cf}

\bibitem[{{Testa} \& {Reale}(2012)}]{Testa2012b}
{Testa}, P., \& {Reale}, F. 2012, \apjl, 750, L10,
  \dodoi{10.1088/2041-8205/750/1/L10}

\bibitem[{{Testa} \& {Reale}(2022)}]{Testa_Reale_2022arXiv}
---. 2022, arXiv e-prints, arXiv:2206.03530.
\newblock \doarXiv{2206.03530}

\bibitem[{{Testa} {et~al.}(2015){Testa}, {Saar}, \& {Drake}}]{Testa2015}
{Testa}, P., {Saar}, S.~H., \& {Drake}, J.~J. 2015, Philosophical Transactions
  of the Royal Society of London Series A, 373, 20140259,
  \dodoi{10.1098/rsta.2014.0259}

\bibitem[{{Testa} {et~al.}(2013){Testa}, {De Pontieu},
  {Mart{\'{\i}}nez-Sykora}, {DeLuca}, {Hansteen}, {Cirtain}, {Winebarger},
  {Golub}, {Kobayashi}, {Korreck}, {Kuzin}, {Walsh}, {DeForest}, {Title}, \&
  {Weber}}]{Testa2013}
{Testa}, P., {De Pontieu}, B., {Mart{\'{\i}}nez-Sykora}, J., {et~al.} 2013,
  \apjl, 770, L1, \dodoi{10.1088/2041-8205/770/1/L1}

\bibitem[{{Testa} {et~al.}(2014){Testa}, {De Pontieu}, {Allred}, {Carlsson},
  {Reale}, {Daw}, {Hansteen}, {Martinez-Sykora}, {Liu}, {DeLuca}, {Golub},
  {McKillop}, {Reeves}, {Saar}, {Tian}, {Lemen}, {Title}, {Boerner},
  {Hurlburt}, {Tarbell}, {Wuelser}, {Kleint}, {Kankelborg}, \&
  {Jaeggli}}]{Testa2014}
{Testa}, P., {De Pontieu}, B., {Allred}, J., {et~al.} 2014, Science, 346,
  1255724, \dodoi{10.1126/science.1255724}

\bibitem[{{Tian} {et~al.}(2014){Tian}, {DeLuca}, {Cranmer}, {De Pontieu},
  {Peter}, {Mart{\'\i}nez-Sykora}, {Golub}, {McKillop}, {Reeves}, {Miralles},
  {McCauley}, {Saar}, {Testa}, {Weber}, {Murphy}, {Lemen}, {Title}, {Boerner},
  {Hurlburt}, {Tarbell}, {Wuelser}, {Kleint}, {Kankelborg}, {Jaeggli},
  {Carlsson}, {Hansteen}, \& {McIntosh}}]{Tian2014}
{Tian}, H., {DeLuca}, E.~E., {Cranmer}, S.~R., {et~al.} 2014, Science, 346,
  1255711, \dodoi{10.1126/science.1255711}

\bibitem[{{Toriumi} \& {Wang}(2019)}]{Toriumi2019}
{Toriumi}, S., \& {Wang}, H. 2019, Living Reviews in Solar Physics, 16, 3,
  \dodoi{10.1007/s41116-019-0019-7}

\bibitem[{{Uitenbroek}(2001)}]{Uitenbroek2001}
{Uitenbroek}, H. 2001, \apj, 557, 389, \dodoi{10.1086/321659}

\bibitem[{{Wright} {et~al.}(2017){Wright}, {Hannah}, {Grefenstette},
  {Glesener}, {Krucker}, {Hudson}, {Smith}, {Marsh}, {White}, \&
  {Kuhar}}]{Wright2017}
{Wright}, P.~J., {Hannah}, I.~G., {Grefenstette}, B.~W., {et~al.} 2017, \apj,
  844, 132, \dodoi{10.3847/1538-4357/aa7a59}

\end{thebibliography}
\bibliographystyle{aasjournal}



\end{document}